# Global Brain Dynamics During Social Exclusion Predict Subsequent Behavioral Conformity

Nick Wasylyshyn[1,2], Brett Hemenway[3], Javier O. Garcia[1,4], Christopher N. Cascio[2], Matthew Brook O'Donnell[2], C. Raymond Bingham[5], Bruce Simons-Morton[6], Jean M. Vettel[1,4,7], Emily B. Falk[2,8,9]

[1]Human Research and Engineering Directorate, U.S. Army Research Laboratory, Aberdeen Proving Ground, MD 21005

[2]Annenberg School for Communication, University of Pennsylvania, Philadelphia, PA 19104

[3]Department of Computer and Information Science, University of Pennsylvania, Philadelphia, PA 19104

[4]Department of Bioengineering, University of Pennsylvania, Philadelphia, PA 19104

[5]University of Michigan Transportation Research Institute, Ann Arbor, MI 48109

[6]Eunice Kennedy Shriver National Institute on Child Health and Human Development, Bethesda, MD 20892

[7]Department of Psychological and Brain Sciences, University of California, Santa Barbara, Santa Barbara, CA 93106

[8]Marketing Department, Wharton School, University of Pennsylvania, University of Pennsylvania, Philadelphia, PA 19104

[9]Department of Psychology, University of Pennsylvania, Philadelphia, PA 19104

Address correspondence to:

Emily Falk
3620 Walnut Street
Philadelphia, PA 19104
Phone: (215) 898-7041
Fax: (215) 898-2024
emily.falk@asc.upenn.edu

**Abstract**

Individuals react differently to social experiences; for example, people who are more sensitive to negative social experiences, such as being excluded, may be more likely to adapt their behavior to fit in with others. We examined whether functional brain connectivity during social exclusion in the fMRI scanner can be used to predict subsequent conformity to peer norms. Adolescent males (N = 57) completed a two-part study on teen driving risk: a social exclusion task (Cyberball) during an fMRI session and a subsequent driving simulator session in which they drove alone and in the presence of a peer who expressed risk-averse or risk-accepting driving norms. We computed the difference in functional connectivity between social exclusion and social inclusion from each node in the brain to nodes in two brain networks, one previously associated with mentalizing (medial prefrontal cortex, temporoparietal junction, precuneus, temporal poles) and another with social pain (anterior cingulate cortex, anterior insula). Using cross-validated machine learning, this measure of *global network connectivity* during exclusion predicts the extent of conformity to peer pressure during driving in the subsequent experimental session. These findings extend our understanding of how global neural dynamics guide social behavior, revealing functional network activity that captures individual differences.

**Introduction**

Social connection is fundamental to human well-being (Pinquart and Sörenson, 2000; Kawachi and Berkman, 2001; Helliwell and Putnam, 2004) and survival (Berkman and Syme, 1978; House et al., 1982; Kawachi et al., 1996), whereas disconnection from social ties negatively impacts emotional and physical health (Holt-Lunstad et al., 2010; Eisenberger and Cole, 2012; Cacioppo and Cacioppo, 2014). Consequently, people work to remain connected to others and avoid social exclusion. When this harmony is disrupted, such as when a member of a group feels excluded, the individual may feel "social pain" (Eisenberger et al., 2003; Eisenberger and Lieberman, 2004; MacDonald and Leary, 2005) and attempt to understand others' thoughts and feelings (i.e., "mentalizing"; Frith & Frith, 2003) in service of reconnecting with others (Maner et al., 2007). One way that people develop and maintain social harmony with those around them is by conforming to others' attitudes and behavior (Cialdani and Goldstein, 2004). Importantly, however, individuals respond differently to social exclusion (Fenigstein, 1979; Nezlek et al., 1997; Zadro et al., 2006; Waldrip, 2007; DeWall et al., 2012; Cascio et al., 2015) and therefore may be differentially disposed to conform in service of maintaining social harmony. In our work, we directly examine whether functional connectivity across the brain as a whole during exclusion can capture and account for individual differences in conformity behavior.

First, we focused on brain regions implicated in social pain (Eisenberger et al., 2003) and mentalizing (Frith and Frith, 2003) processes, which might take on more importance globally in the brain during exclusion, relative to inclusion, for individuals who are most reactive to the exclusion experience (Eisenberger et al., 2003; Masten et al., 2010). Previous research demonstrates the high cost of social exclusion (see Williams, 2007 for a review) and the fact that neural reactivity to exclusion varies across individuals (Falk et al., 2014). We focused on connectivity for its temporal sensitivity to processes that may be recurrent but not constant, such as thoughts about others' transient mental states (Schmälzle et al., in press). If social pain and mentalizing have more global importance in interpreting experiences during exclusion for some individuals, we would expect not only univariate changes (for a review, see Eisenberger, 2015) but also changes in global connectivity between regions associated with social pain and mentalizing and the rest of the brain. Indeed, a growing body of research demonstrates that large-scale interactions between key regions of interest and the rest of the brain may provide additional information, complementing regional activity in capturing current mental states (van den Heuvel and Hulshoff, 2010; Bassett et al., 2015; Medaglia et al., 2015). Yet, despite recent studies that have begun to consider functional connectivity among single regions during social tasks (Bolling et al., 2011; Meyer, 2012; Puetz et al., 2014), little is known about larger-scale network dynamics during social experiences.

We hypothesized that the extent to which people conform to their peers can be predicted using changes in brain dynamics linked to social pain and mentalizing when people are faced with social exclusion (cf., Falk et al., 2014). Although there is promise of predicting behavior from network dynamics (Bassett et al., 2011; Baldassarre et al., 2012), research has not yet linked brain network dynamics during social tasks in the neuroimaging environment to objectively logged social behaviors measured outside of the scanner. To this end, we examined the relationship between global connectivity of regions implicated in social pain and mentalizing during social exclusion and inclusion as predictors of conformity to peer influence on simulated driving outside of the scanner a week later. We focused on brain dynamics during exclusion given past research demonstrating that conformity is one way that participants try to regain acceptance (DeWall, 2010) and potentially preempt further exclusion. Conformity to driving risk attitudes in teens served as our outcome because teens' driving risk is socially influenced (Simons-Morton et al., 2005; Simons-Morton et al., 2014; Bingham et al., 2016) and has important real-world consequences (Ouimet et al., 2010).

## Materials and Methods

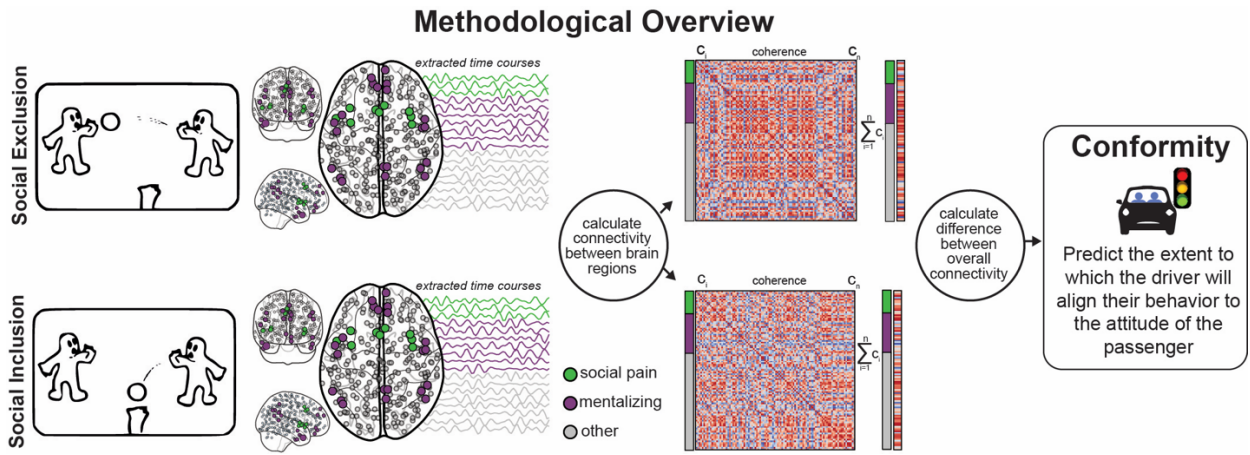

Figure 1. Study Overview. fMRI BOLD data were collected during Cyberball, a virtual ball-tossing game that simulates social exclusion and social inclusion. Functional brain activity was extracted from regions in a whole-brain parcellation (gray), including regions previously associated with social pain (green) and mentalizing (purple). Connectivity was then computed between all region pairs, and the difference in connectivity during social exclusion and social inclusion was used to predict subsequent conformity to the attitude of a peer passenger during a driving simulator.

This research was part of a larger investigation of the effects of peer influence on teen driving (Cascio et al., 2014; Falk et al., 2014; Simons-Morton et al., 2014; Bingham et al., 2016; Schmälzle et al., in press). Prior reports on the driving simulator data reported here, examining the effects of peer influence on teen driving behavior, noted substantial individual variability in susceptibility to influence (Bingham et al., 2016). In this analysis, we extend these results by examining a new measure of global functional connectivity during exclusion vs. inclusion in the brain during Cyberball and investigate its predictive relationship with individual differences in conformity (Figure 1).

*Participants*

Fifty-seven right-handed, neurotypical, male participants aged 16 or 17 completed both portions of the study. Each participant had received a Level 2 Michigan driver's license (unsupervised driving allowed with several restrictions) at least 4 months prior to the study, drove at least twice per week, had normal or corrected-to-normal vision, and was insensitive to simulator sickness. Participants were told that the purpose of the study was to examine the physiology of driving. The University of Michigan Institutional Review Board approved the study procedures; participants provided written assent, and their legal guardians provided written informed consent.

*Neuroimaging Data Collection: Cyberball (fMRI)*

Participants were told that they would play a variety of computer games while in a functional magnetic resonance imaging (fMRI) scanner, some alone and one, called Cyberball, with two other participants; the other "participants" were in reality controlled by a computer (Williams et al., 2000; Williams and Jarvis, 2006). Participants played two rounds of Cyberball. The first condition simulated social inclusion by having each player (the actual participant and the two simulated players) receive the ball equally often; in the second condition, the game started the same as the first but the two computer-controlled players soon began throwing the ball only between each other, excluding the participant. Both the inclusion and exclusion conditions lasted approximately 2.5 minutes (74 brain volumes for each condition).

Functional images were recorded using a reverse spiral sequence (TR = 2000 msec, echo time = 30 msec, flip angle = 90°, 43 axial slices, field of view = 220 mm, slice thickness = 3 mm, voxel size = 3.44 × 3.44 × 3.0 mm).

*Need-Threat Scale: Measuring the Effects of Ostracism*

Following the fMRI scan, participants completed several questionnaires, including the Need-Threat Scale (van Beest and Williams, 2006). This assessment quantifies the perceived threat to participants' social needs experienced during Cyberball on a scale of 1 to 7, with lower scores indicating higher threat. Scores ranged from 1.7 to 6.1, with a mean of 3.48 and a standard deviation of 0.96.

*Neuroimaging Data Analysis: Global Connectivity*

Following a standard preprocessing stream (see Supplementary Material), we used a previously published whole-brain parcellation (Power et al., 2011) to define 264 regions. From this atlas, we then identified the regions that were closest to the centroid of a set of brain areas associated with social pain, including the anterior cingulate cortex (ACC) and anterior insula (AI) (Eisenberger, 2003; Eisenberger and Lieberman, 2004; Lamm and Singer, 2010; Cacioppo et al., 2013; Rotge et al., 2014), and a separate set involved in mentalizing, including the temporoparietal junction (TPJ), temporal pole (TP), precuneus (PC), and dorsomedial and ventromedial prefrontal cortex (PFC) (Frith and Frith, 2003; Frith and Frith, 2006; D'Argembeau et al., 2007; Van Overwalle and Baetens, 2009).

Since the regions of these two networks are subdivided in the whole-brain atlas used (Power et al., 2011), we selected the three nodes in the parcellation that were closest to the center of each of the 10 theory-driven regions (D'Argembeau et al., 2007; Schmälzle et al., in press; coordinates in Figure S1). This resulted in 30 nodes to represent the two theory-driven networks from this atlas (depicted by purple and green nodes for mentalizing and social pain, respectively, in Figure 1).

To estimate functional connectivity between brain regions during the social inclusion and exclusion conditions of the Cyberball game, we calculated the coherence (Rosenberg et al., 1989) between every pair of regions (see Supplemental Material for details).

For each region, we derived a measure of "global connectivity" for the social inclusion condition and one for the social exclusion condition. We created a graph with 264 nodes, one for each region in the whole-brain atlas (Power et al., 2011), and the edge weight between nodes was determined by their coherence. The weighted degree of a region is the sum of its coherence to every other region in the brain and represents the extent to which the region is connected to the rest of the brain. We used this weighted degree summation as the global connectivity metric for each region, computing it separately for the two Cyberball conditions. We then subtracted the weighted degree during the inclusion condition from the weighted degree during the exclusion condition. The difference in a node's weighted degree encapsulates the effect that social exclusion has on the region's global connectivity (Figure 1). We then use the global connectivity metric for each region as the feature set in our machine learning analysis to predict a participant's behavioral conformity in a subsequent driving simulator session.

*Behavioral Data Collection: Driving Simulator*

Approximately one week after the fMRI session, participants returned for the driving simulator session. After a short practice drive, participants completed a solo practice drive in the simulator and then two drives, one solo and one with a confederate as a passenger. Order of the drive conditions was counterbalanced between participants. Each drive lasted 10-15 minutes, and the participant approached either 9 or 10 traffic lights that were timed to turn red before the participant cleared the intersection. The timing of the yellow lights was fine-tuned after the first 8 participants and was held constant for the final 49 participants. To validate that the timing changes did not influence the results, we performed a two-sided t-test to compare conformity between the first 8 participants and the final participants and determined that the two samples were not significantly different; t(55) = 1.33, p = 0.190. Additional information about the driving simulator can be found in the Supplementary Material.

Each participant was randomly assigned (between participants) to one of two conditions for the passenger drive: risk-averse or risk-accepting passenger. In both conditions, participants completed a pre-drive survey after which a similar-aged, male confederate arrived late. In the risk-averse condition, he explained, "Sorry I was a little late getting here. I tend to drive slower, plus I hit every yellow light," whereas the risk-accepting confederate said, "Sorry I was a little late getting here. Normally I drive way faster, but I hit like every red light."

In addition, before the passenger drive, the participant and confederate watched two videos together, one showing high-risk driving from the passenger seat and one showing low-risk driving from the same perspective; the order in which the videos were shown was random. After each video, the participant and confederate were asked to rate on a scale of 1 to 10 how similar their driving was to the driving shown in the video and how likely they would be to ride with the driver in the video. The confederate responded second and gave responses that were less risky than the participant's in the risk-averse condition or riskier than the participant's in the risk-accepting condition.

Finally, during the passenger drive, the confederate was told to navigate using a map, which provided an excuse to make comments about the participant's driving behavior during the drive. In the risk-accepting condition, the confederate stated the speed limit when the participant was driving below it, while in the risk-averse condition, he noted reduced-speed zones.

*Behavioral Data Analysis: Conformity*

For each drive, we derived a measure of conformity by calculating the percentage of intersections where the driver failed to stop in each drive and subtracting them, such that a positive difference in either condition indicates that the participant conformed to the confederate's attitude by driving in a riskier manner in the risk-accepting condition and more safely in the risk-averse condition. This estimation of conformity serves as the dependent variable for our analysis.

*Cross-validated Analysis: Connectivity Predicts Conformity*

We used a machine learning pipeline to examine whether brain connectivity predicts behavioral conformity in an out-of-scanner driving task a week later, using global connectivity of regions as features. The primary analysis used 30 regions from the theory-driven brain areas as features; a second (see Supplementary Material) used all 264 from the whole-brain parcellation. To prevent overfitting, we created 57 splits into training and test sets, each one leaving out one participant from the training set, and performed feature selection within each split, resulting in a model with 7 features (see Supplementary Material for details on feature selection).

**Results**

We investigated whether functional connectivity during exclusion relative to inclusion in the Cyberball task in the scanner predicted conformity to a confederate passenger in a simulated drive one week later. Our primary analysis examined connectivity of theory-driven regions involved in mentalizing and social pain. Functional connectivity was computed between each region and the rest of the brain, and these metrics of global connectivity were used as features in a cross-validated machine learning analysis to predict conformity in the driving simulator.

*Behavioral conformity responses in the driving simulator*

We first examined drivers' conformity to the attitude of their confederates defined as the amount to which they moved their driving behavior in the direction of the confederates during the passenger drive relative to the solo drive, i.e., increased risk in the presence of a passenger in the risk-accepting condition and decreased risk in the presence of a passenger in the risk-averse condition. Twenty-six drivers were randomized to the risk-accepting condition, and 31 were in the risk-averse condition. The majority of participants (70%) conformed to the confederate's disposition (n = 26) or drove equally safely in each trial (n = 14), although a subset of participants (n = 17) altered their behavior in the opposite direction of the confederate (e.g., drove more riskily in the presence of a risk-averse confederate). The distribution of conformity in the risk-accepting (M = 1.90% change in risk toward the passenger's attitude, SD = 24.65%) and risk-averse conditions (M = 1.92%, SD = 25.14%) was nearly identical. The combined sample had a mean of 1.91% change toward the confederate's attitude and a standard deviation of 24.65%. Consistent with behavioral reports using these data (Bingham et al., 2016), these behavioral results indicate a slight bias toward conformity to the confederate's attitude, but the large amount of variance confirms its value for understanding individual differences in susceptibility to social influence for risky behaviors.

*Global Connectivity from Theory-Driven Regions to the Rest of the Brain Predicts Subsequent Conformity*

We then investigated whether individual differences in the global connectivity of social pain and mentalizing regions (Figure 1, green and purple regions, respectively) during Cyberball could account for the substantial amount of variability in conformity found in the subsequent driving session. We used global connectivity from the 30 theory-driven regions as features in a leave-one-out cross-validation to predict a participant's conformity score in the driving simulator one week later. A parallel analysis conducted considering all 264 regions in the whole-brain parcellation (Power et al., 2011) confirmed the importance of these networks (see Supplementary Material).

To assess global connectivity, we first computed the difference in a region's global connectivity with the rest of the brain between the two social conditions of Cyberball (social exclusion - social inclusion). We then used these global connectivity measures for each person within our theory-driven regions of interest to predict individual differences in conformity during the driving session. As shown in Figure 2 (top left), the best prediction was achieved from 7 of the 30 theory-driven regions; these 7 regions' connectivity predicted individual differences in conformity with an out-of-sample $R^2$ of 0.325 (root mean squared error of 20.07 in cross-validation). Two of the 7 regions selected were in the social pain network, and the other 5 were from the mentalizing network. In Figure 2, the regions in the social pain network are outlined in green, and the regions in the mentalizing network are outlined in purple; the center color for each region reflects the regression coefficient to further characterize the predictive relationship between global connectivity from each region and subsequent conformity. These regions include two regions in the right TPJ, one in the left TPJ, one in the left TP, and bilateral regions in the AI. For 5 of the 7 regions selected, more connectivity during social exclusion than inclusion was predictive of behavioral conformity.

To verify that this result does not arise from chance or overfitting, we performed a permutation test on the data, training our model on 10,000 permutations of the dependent variable and using it to predict the shuffled dependent variables in leave-one-out cross-validation. The true model achieved a better $R^2$ than 99.99% of the shuffled models (p = 0.0002; Ojala and Garriga, 2010). As a final test of significance, we generated 1000 random subsets of 30 regions from the entire 264-region whole-brain atlas (Power et al., 2011) and ran our model starting with these regions instead of our 30 theory-driven regions. For each random subset of 30 regions, we tested models using between 1 and 10 features and empirically selected the optimal number of features for this subset of regions. This led to 1000 optimized models. Only 0.7% of these optimized models outperformed our original model, with more than two thirds of the total features selected in those models coming from among the features selected from our networks-of-interest or whole-brain models (see Supplemental Material). The median out-of-sample $R^2$ score among all 1000 optimized models was 0.055.

Finally, we examined whether global connectivity during either of the two Cyberball conditions could predict conformity as well as did their difference. Individual differences in connectivity during social exclusion were substantially predictive of conformity ($R^2$ = 0.274), although not as strongly as the difference in connectivity between social exclusion and social inclusion ($R^2$ = 0.325); individual differences in connectivity during social inclusion had little predictive power ($R^2$ = 0.096).

Collectively, these results indicate that global connectivity during social exclusion, either alone or in comparison to connectivity during social inclusion, can predict individual differences in subsequent conformity behavior one week later. This highlights the value of examining global connectivity to understand individual variability in real-world social situations.

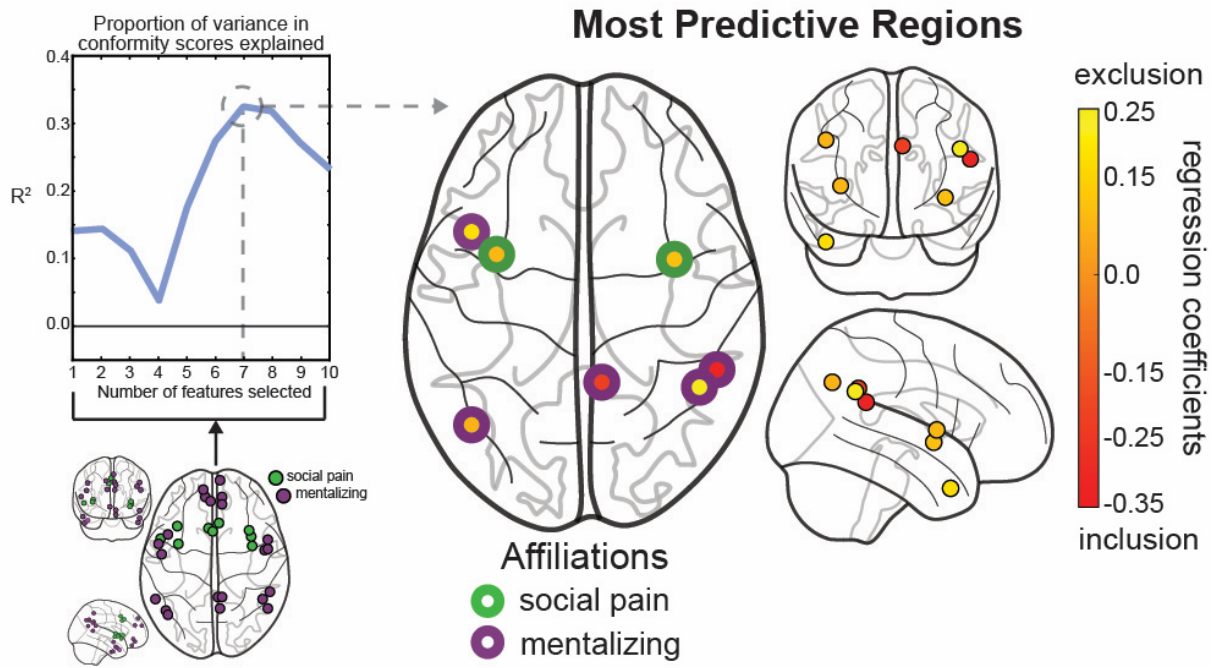

Figure 2. The regions in the social pain and mentalizing networks whose connectivity to the rest of the brain is most predictive of conformity; regions pictured in green and purple, respectively, in bottom left. Top left: The out-of-sample R2 values obtained when selecting between 1 and 10 regions as features in our model, with 7 features being most predictive of conformity. Right: The 7 regions selected by our feature-selection algorithm. The regions with a green outline are in the social pain network, while the regions with a purple outline are in the mentalizing network. The color of the center of each region indicates its regression coefficient. For yellow regions, more connectivity during social exclusion than social inclusion is predictive of conformity to the confederate's attitude; for red regions, the opposite is true.

**Discussion**

Social connection is fundamental to well-being, and a motivating force for a wide range of behaviors, including conformity (Cialdani and Goldstein, 2004; Maner et al., 2007). Previous research has characterized a neural alarm system that responds to social pain (Eisenberger, 2003; Eisenberger and Lieberman, 2004; Lamm and Singer, 2010; Rotge et al., 2014), as well as a broader set of brain regions that allow people to understand others' mental states (Frith and Frith, 2003; Frith and Frith, 2006; D'Argembeau et al., 2007; Van Overwalle and Baetens, 2009). Using a Cyberball game, we show that individual differences in the degree to which key brain regions implicated in social pain and mentalizing change their connectivity with the rest of the brain in response to social exclusion predict conformity to peer attitudes in a driving simulator a week later. Thus, the current research uses a novel network neuroscience perspective to highlight how individual differences in network connectivity in response to a social experience, such as exclusion, predict sensitivity to social influence in a real-world setting.

Although we observed non-significant trends in group-averaged neural responses to social exclusion, not all participants showed equal levels of differentiation in their global connectivity between exclusion and inclusion. Similarly, in the initial report of the driving simulator data used here, the authors noted substantial individual variability in tendency to conform (Bingham et al., 2016). Previous research suggested that sensitivity to social pain might prime individuals to preempt exclusion in other social contexts by conforming (Maner et al., 2007; Peake et al., 2013; Falk et al., 2014). For example, Falk and colleagues (2014) found that univariate increases in brain activity within social pain and mentalizing regions of interest interest were associated with greater driving risk taking in the presence of a peer, compared to driving alone. No prior research, however, has examined how social pain and mentalizing regions might change their global connectivity with the rest of the brain in response to social threats, nor how this relates to real-world relevant decision-making. In this study, we hypothesized that individual differences in connectivity between social pain and mentalizing systems with the rest of the brain would relate to behavioral conformity responses in an unrelated driving context. Consistent with this hypothesis, coherence in brain networks involved in responding to social cues (i.e., social pain and mentalizing networks) during social exclusion compared to social inclusion predicted approximately one third of the variance in the degree to which participants conformed to peers' driving preferences a week later. This result substantially extends past research on social behavior and the brain by demonstrating that the global connectivity of social pain and mentalizing systems in response to exclusion maps onto the inclination to conform to peer attitudes. These data are consistent with the idea that conformity is a means to preserve one's position in a group and that a person who experiences a greater reaction to exclusion may take greater actions to prevent such an experience in other contexts.

The current findings also extend past research by revealing information about how the brain helps navigate the social world. In line with past research suggesting that specific control points in the brain, and particularly within the default mode network, help transition the brain to execute different tasks (Gu et al., 2015; Betzel et al., 2016; Medaglia et al., submitted), we find that global connectivity between key social pain and mentalizing regions predicts individual differences in susceptibility to peer influence. In other words, greater changes in global brain connectivity may be associated with flexibly altering behaviors to adjust to social situations. Specifically, our method used functional connectivity in a novel way that allowed us to identify specific regions whose global brain dynamics were the most predictive of behavior change. During our main analysis, these regions were selected from two hypothesized networks of interest, namely networks previously associated with social pain and mentalizing. Prior work has shown that activity in the social pain and mentalizing networks can be used to predict subsequent behavior change (Hein et al., 2010; Carter et al., 2012). Here, we show that functional connections between both regions in the social pain (e.g., bilateral AI) and mentalizing networks (bilateral TPJ), and the rest of the brain are associated with later individual differences in tendency to change behavior. Both the right TPJ and right AI also appeared in our whole-brain analysis (see Supplemental Material), suggesting the robustness of these results.

The key roles of the AI and TPJ in our models may elucidate the psychological significance of our method. The right AI has been identified as a "causal outflow hub" (Sridharan et al., 2008; Menon and Uddin, 2010; Uddin et al., 2011), meaning that its activity is predictive of that of a large number of other regions in the brain. Similarly, the TPJ also functions as a hub of connectivity, integrating activity in different regions into a single coherent social context and affecting processing throughout the brain (Carter and Huettel, 2013). As the features in our model are the cumulative (i.e., "global") functional connectivity of each region, it is to be expected that the regions that are most predictive of behavior change are those that serve as focal points, integrating and influencing other regions.

Taken together, these results highlight the importance of considering not only how individual brain regions are modulated by social experiences but also how those regions communicate with the rest of the brain more globally. We find that social context (i.e., exclusion vs. inclusion) causes different changes across individuals in the extent to which key regions implicated in social pain and mentalizing become more globally connected to the rest of the brain. Further, individual differences in the extent of this shift were significantly predictive of later conformity to driving norms expressed by a peer.

*Future Directions*

Our results show a predictive relationship between brain activity and social influence in our sample of 16- and 17-year-old, primarily Caucasian, males from Southeast Michigan. Future research could examine whether this relationship changes across developmental periods, including whether the brain regions involved in responding to social exclusion fluctuate over time or play a differential role in the brain's global connectivity dependent on developmental stage (see Vijayakumar et al., 2017 for a univariate perspective on this question). It would also be interesting to examine whether this relationship generalizes across other socio-demographic populations since cultural variation has been shown to influence social processing, including social orientation (individualism versus collectivism; Markus and Kitayama, 1991), decision-making (Iyengar and Lepper, 1999), and team performance (Wagner et al., 2012).

*Conclusion*

This work shows that the functional connectivity of brain regions associated with social pain and mentalizing in response to social exclusion is able to predict subsequent conformity. This result highlights the power of considering global connectivity as predictor and is a first step toward understanding how neural connectivity informs our interaction with the social world. The technique that we developed in the process, using overall connectivity of regions as predictors, addresses common limitations of other connectivity techniques while capturing processes that are averaged away in models based on mean activation. Our method is likely to have applications for developing predictive models based on network dynamics, which in turn provide parsimonious explanations relating brain activity, social context and behavior.

**Funding**

The research was supported by (1) the intramural research program of the *Eunice Kennedy Shriver* National Institute of Child Health and Human Development contract # HHSN275201000007C (PI:Bingham); (2) A University of Michigan Injury Center Pilot Grant (PI:Falk); (3) #NIH/NICHD IR21HD073549- 01A1 (PI:Falk); (4) An NIH Director's New Innovator Award #1DP2DA03515601 (PI:Falk), and (5) U.S. Army Research Laboratory, including work under Cooperative Agreement W911NF-10-2-0022 and W911NF-16-2-0165.

**Acknowledgements**

The authors gratefully acknowledge the Communication Neuroscience lab for research assistance and the staff of the University of Michigan fMRI Center as well as Jean T. Shope, Marie Claude Ouimet, Anuj K. Pradhan, Kristin Shumaker, Jennifer LaRose, Farideh Almani, and Johanna Dolle for collaboration on a larger study from which these data were drawn and assistance with data collection. The authors gratefully acknowledge Andrew Suzuki, Robin Liu, Ryan Bondy, Matthew Sweet, Cary Welsh, Andrea I. Barretto, Jennifer LaRose, Farideh Almani, Alyssa Templar and Kristin Shumaker for research assistance, and the staff of the University of Michigan fMRI Center. We thank Ralf Schmälzle for provision of ROIs.

**Conflict of Interest Statement**

The authors have no conflicts of interest to declare.

**Supplementary Material**

*Preprocessing Stream*

To enhance coregistration and normalization, in-plane T1-weighted images (43 slices, slice thickness = 3 mm, voxel size = 0.86 × 0.86 × 3.0 mm) and high-resolution T1-weighted images (SPGR, 124 slices, slice thickness = 1.02 × 1.02 × 1.2 mm) were also acquired. After discarding the first four acquired volumes, the functional data were preprocessed and analyzed using Statistical Parametric Mapping (SPM8, Wellcome Department of Cognitive Neurology, Institute of Neurology, London, UK), and images despiked using the 3dDespike program as implemented in the AFNI toolbox (Cox, 1996; Cox and Hyde, 1997; Gold et al., 1998). The volumes were then corrected for slice time acquisition differences and spatially realigned to the first functional image. Functional and structural images were coregistered by aligning the in-plane T1 images to the mean functional image, and then the in-plane image was registered to the high-resolution T1 images. Structural images were then skull-stripped and normalized to the skull-stripped MNI template provided by FSL.

*Driving Simulator and Environment Specifications*

The driving simulator was manufactured by DriveSafety and included the front three quarters of the body and the front interior of a sedan. It features a projected LCD instrument cluster controlled by a computer, foot controls, and realistic steering force feedback. The virtual road environment was projected onto three forward screens and one rear screen, each with a resolution of 1024 x 768 pixels, for 120 degrees of forward view and 40 degrees of rear view visible through the side and rearview mirrors. The driving environment contained standard roads, intersections, traffic control devices, other vehicles, and pedestrians; in addition, purely visual elements including buildings, trees, and sky were present. The simulator's sound system produced realistic interior and exterior sounds, and road vibration was simulated through the floorboard.

*Coherence Calculation*

An advantage of coherence compared to correlation in connectivity analysis is its relative insensitivity to differences in the shape of the hemodynamic response function (HRF) (Sun et al., 2004; Lauritzen et al., 2009), which has been shown to vary between regions in the same individual's brain (Handwerker et al., 2004). Coherence was computed using Welch's method with a 48-point discrete Fourier transform Hanning window and a 24-point overlap between windows (Welch, 1967). In accordance with Lauritzen et al. (2009), we analyzed coherence for 9 frequency bands centered at 0.0625, 0.0729, 0.0833, 0.0938, 0.1042, 0.1146, 0.125, 0.1354, and 0.1458 Hz; we analyzed each frequency band individually instead of averaging over them. The most predictive relationships were found using a frequency of 0.1146 Hz, so all results are presented from this band.

*Networks of Interest*

We utilized the same centers for the regions of interest in the theory-driven analysis (social pain and mentalizing networks) as in Schmälzle et al. (in press), with the exception of replacing the nodes in the lateral temporal cortex with the temporal poles by choosing the closest regions in the atlas (Power et al., 2011). This change was made in order to be more consistent with the literature, especially D'Argembeau et al. (2007), from which the coordinates are drawn, and does not substantively change the conclusions reported. The regions are pictured below along with their MNI coordinates.

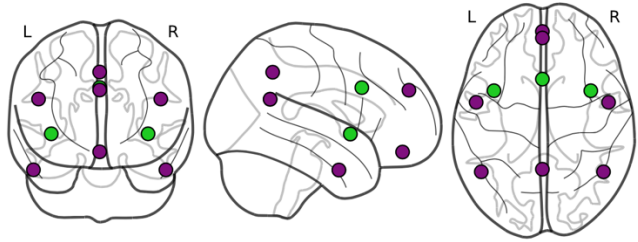

Figure S1. The locations of the social pain and mentalizing networks' regions of interest. The mentalizing network is shown in purple, and the social pain network is green.

| Network | Region | Coordinates |
| --- | --- | --- |
| Social pain | ACC | (0, 16, 32) |
| Social pain | left AI | (-38, 7, -4) |
| Social pain | right AI | (38, 7, -4) |
| Mentalizing | dorsomedial PFC | (0, 53, 30) |
| Mentalizing | ventromedial PFC | (0, 48, -18) |
| Mentalizing | PC | (0, -54, 44) |
| Mentalizing | left TPJ | (-48, -56, 23) |
| Mentalizing | right TPJ | (48, -56, 23) |
| Mentalizing | left TP | (-52, -2, -32) |
| Mentalizing | right TP | (52, -2, -32) |

Table S1. MNI coordinates of the centroids of the regions in the social pain and mentalizing networks of interest.

*Global Connectivity and Overfitting*

Our method relies on global connectivity for each region to minimize the risk of overfitting a predictive model from connectivity data. One danger when using connectivity as a predictor is that a network with n nodes has n-choose-2 potential pairwise connections. Thus, a model based on individual activity has n possible predictors, whereas a model taking into account all possible pairwise connections has almost $n^2/2$ possible predictors; it is easy to overfit such a model. To mitigate this risk, our method calculates the strength of each region's connection to the rest of the brain instead of building a model incorporating the pairwise coherence between each region. This technique limits the number of potential predictors to the number of regions, thus diminishing the possibility of overfitting compared to a model examining individual pairwise connections.

*Feature Selection in Scikit-Learn*

For each training set of 56 participants, we used scikit-learn's SelectKBest algorithm to sequentially select the k best features according to their F scores (Pedregosa et al., 2011). Common wisdom suggests choosing approximately one feature for every 10-15 observations for small sample sizes (Babyak, 2004), corresponding to between 4 and 6 features for our dataset; however, we expanded our search to 10 since our features were theoretically motivated. We then performed leave-one-out cross-validation using ordinary least squares regression so that each participant's conformity to peer influence was predicted using a model trained only on the other 56 participants. In the first analysis with 30 theory-driven regions, the model with highest accuracy in cross-validation occurred with $k = 7$, so we focused on the 7-feature model. In the second whole-brain analysis with 264 regions, models with 5 and 6 features performed very well, and we opted to select 6 features based on cross-validation scores.

*Controlling for Drive Order*

Because participants completed the solo and passenger drives in random order, we performed a set of cross-validated analyses to account for the possibility of drive order as a confounding variable. Specifically, we trained a model with 31 potential features: the 30 brain regions from the networks of interest plus drive order. However, drive order was never selected by SelectKBest, so we did not pursue this approach further.

*Connectivity Differences in Response to Social Exclusion*

Although the mean connectivity to the rest of the brain in seven of the nine regions in the social pain network was higher during exclusion than during inclusion, none was significantly different

(Figure S2). Similarly, we observed nonsignificant trends in the mentalizing network: Five of the six regions in the medial PFC were more connected to the whole brain on average during inclusion than during exclusion, whereas each region in the bilateral temporal poles exhibited more global connectivity during exclusion than during inclusion on average. Overall, although we observed weak connectivity differences between Cyberball conditions, with two thirds of the regions more strongly connected to the rest of the brain during exclusion than during inclusion, these differences were not significant or robust. This arose due to substantial heterogeneity across participants in the degree of change from exclusion to inclusion, highlighting the importance of considering individual differences.

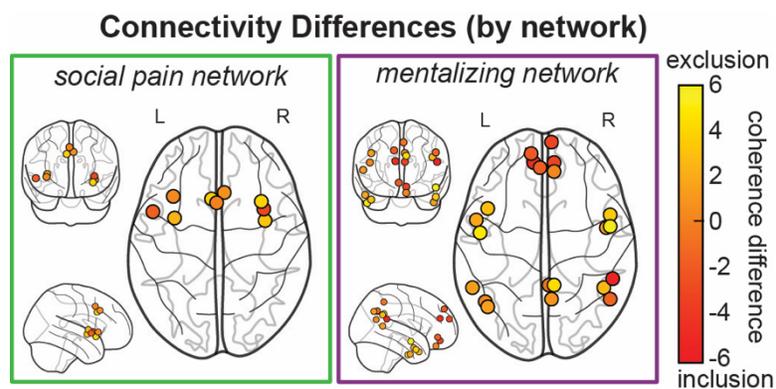

Figure S2. Differences in global connectivity between social exclusion and social inclusion in the social pain network (left) and the mentalizing network (right), averaged across all participants. The nodes with more connectivity to the rest of the brain during exclusion than during inclusion are yellow, and the nodes that are more connected during inclusion than during exclusion are red. Note: none of the differences pictured is statistically significant; results are depicted for descriptive purposes.

*Whole-brain Connectivity Analysis Confirms Importance of Theory-Driven Regions*

We replicated our analysis using theory-driven networks using a whole-brain analysis. This analysis was conducted in part to assess the robustness of the first analysis, investigating whether the *a priori* networks of interest would still be identified among the 264 whole-brain regions. Furthermore, it examined whether the out-of-sample $R^2$ value could be substantially improved when the connectivity features were not limited to approximately ten percent of the possible regions (i.e., when not limiting to theory-driven ROIs but rather searching across the whole brain).

As shown in Figure S3 (top left), the best prediction was achieved from 6 of the 264 whole-brain regions, and these top 6 regions predicted individual differences in conformity with an out-of-sample $R^2$ of 0.303 (root mean squared error of 20.40 in cross-validation), which is slightly less but comparable to the theory-driven analysis. Half of these regions overlap with or neighbor the theory-driven social pain and mentalizing regions, including one region in the AI (social pain region, outlined in green in Figure S3) and two regions in the right TPJ (one mentalizing region, outlined in purple in Figure S3, and an adjacent TPJ atlas ROI). Furthermore, the whole-brain analysis identified three additional regions, two in the left motor/premotor cortex and one in the right dorsolateral PFC (outlined in black in Figure S3).

Following the same procedure used in the theory-driven analysis, we performed 10,000 permutations to verify that this result does not arise from chance or overfitting, and the whole-brain model outperformed 99.77% of the shuffled models (p = 0.0024; Ojala and Garriga, 2010). We also confirmed that the difference in connectivity between exclusion and inclusion was more predictive than that of either state alone, with coherence calculated at any frequency band, with social exclusion being more predictive ($R^2$ = 0.219) than social inclusion ($R^2$ = 0.161).

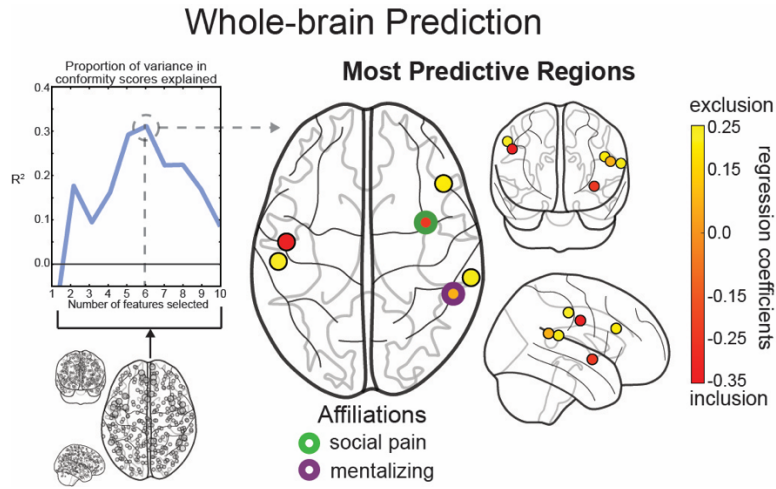

Figure S3. The regions from the whole brain atlas whose connectivity to the rest of the brain is most predictive of conformity. Top left: The out-of-sample $R^2$ values obtained when selecting between 1 and 10 regions as features in our model, with 6 features being most predictive of conformity. Right: The 6 regions selected by our feature-selection algorithm. The region with a green outline is in the social pain network, and the region with a purple outline is in the mentalizing network. The other four regions were not among the 30 regions representing either network of interest. The color of the center of each region indicates its regression coefficient. Regions for which more connectivity during social exclusion than social inclusion is predictive of conformity are yellow, and regions for which more connectivity during social inclusion than social exclusion is predictive of conformity are red.